\documentclass[9pt,letterpaper,technote]{IEEEtran}

\usepackage{amssymb}
\usepackage{amsmath}
\usepackage{amsfonts}
\usepackage{graphicx}
\usepackage{color}
\usepackage{dcolumn}
\usepackage{bm}
\newtheorem{definition}{Definition}
\newtheorem{theorem}{Theorem}
\newtheorem{lemma}{Lemma}

\def\aa{\vec{\alpha}}
\def\vec#1{{\bm #1}}

\def\ket#1{| #1 \rangle}

\def\norm#1{\| #1 \|}
\def\d{\partial}
\def\diag{\operatorname{diag}}

\def\dim{\operatorname{dim}}
\def\supp{\operatorname{supp}}

\def\Tr{\operatorname{Tr}}

\def\H{\mathcal{H}}
\def\X{\mathcal{X}}

\def\D{\mathcal{D}}

\def\N{\mathcal{N}}

\def\CC{\mathbb{C}}
\def\RR{\mathbb{R}}

\def\BB{\mathfrak{B}}
\def\DD{\mathfrak{D}}
\def\HH{\mathfrak{H}}

\def\SGS#1{\color{black} #1 \color{black}}

\newcommand{\tr}{\textrm{tr}}
\newcommand{\trace}{\textrm{trace}}

\newcommand{\beq}{\begin{equation}}
\newcommand{\eeq}{\end{equation}}
\newcommand{\beqa}{\begin{eqnarray}}
\newcommand{\eeqa}{\end{eqnarray}}
\newcommand{\beqan}{\begin{eqnarray*}}
\newcommand{\eeqan}{\end{eqnarray*}}
\newcommand{\bea}{\begin{eqnarray}}
\newcommand{\eea}{\end{eqnarray}}

\begin{document}

\title{Stabilizing Quantum States by Constructive Design of 
Open Quantum Dynamics}

\author{Francesco Ticozzi, Sophie G. Schirmer and Xiaoting Wang
\thanks{F. Ticozzi is with the Dipartimento di Ingegneria
dell'Informazione, Universit\`a di Padova, via Gradenigo 6/B, 35131
Padova, Italy ({\tt ticozzi@dei.unipd.it}).} 
\thanks{S.~G.~Schirmer and X.~Wang are with the Dept of Applied Maths
\& Theoretical Physics, University of Cambridge, Wilberforce Rd,
Cambridge, CB3 0WA, United Kingdom ({\tt sgs29@cam.ac.uk}, {\tt
xw233@cam.ac.uk}).}
\thanks{F.T. acknowledges support by the CPDA080209/08 and ``Quantum
Future'' research grants of the University of Padova, and by the
Department of Information Engineering research project ``QUINTET''.  SGS
acknowledges funding from EPSRC ARF Grant EP/D07192X/1 and Hitachi.} 
}
\date{\today}

\maketitle

\begin{abstract}
Based on recent work on the asymptotic behavior of controlled quantum
Markovian dynamics, we show that any generic quantum state can be
stabilized by devising constructively a simple Lindblad-GKS generator
that can achieve global asymptotic stability at the desired state. 
The applicability of such result is demonstrated by designing a direct
feedback strategy that achieves global stabilization of a qubit state
encoded in a noise-protected subspace.
\end{abstract}

\section{Introduction}

The theory of open quantum dynamics has attracted significant interest
recently due to the fast development of new experimental skills to
study, and even design, the interaction between a quantum system and its
environment. In many applications, the dynamics of an open system
interacting with its reservoir can be described by a quantum Markov
process. Specifically, let us consider a finite-dimensional quantum
system described in a Hilbert space $\H\simeq\CC^N.$ The state of the
system is represented by a \emph{density operator} $\rho$ on $\H$ with
$\rho\geq 0$ and $\trace(\rho)=1$.  Density operators form a convex set
$\DD(\H)\subset\HH(\H)$ with one-dimensional projectors corresponding to
extreme points (\emph{pure states}).  We denote by $\BB(\H)$ the set of
linear operators on $\H$, with $\HH(\H)$ denoting the real subspace of
Hermitian operators. Throughout the paper we will use $\dag$ to denote
the adjoint, $*$ for the complex conjugate, and $[X,Y]=XY-YX,$
$\{X,Y\}=XY+YX$ to represent the commutator and the anticommutator,
respectively.  The Markovian quantum dynamics is described by the
following Lindblad-Gorini-Kossakowskii-Sudarshan master equation
(LME)~\cite{lindblad,gorini-k-s}:
\begin{equation}
\label{gen}
 \dot\rho = -i[H,\rho]+\D(L,\rho)
          = -i[H,\rho]+L\rho L^\dag -\tfrac{1}{2}\{L^\dag L,\rho\},
\end{equation}
where $H\in\HH(\H)$ is the Hamiltonian and $L\in\BB(\H)$ describes the
dissipation due to the environment, accounting for the non-unitary part
of the evolution.

The mathematical problem we consider is the following: \emph{Find a pair
$(H,L)$ that makes a given density matrix $\rho$ globally asymptotically
stable (GAS) assuming dissipative dynamics of Lindblad type} (for the
standard definition of GAS see~\cite{khalil}).  This type of dynamic
stabilization of quantum states is important in quantum information
processing applications~\cite{viola-qubit,nielsen-chuang}.  The main
result of this paper is a constructive procedure to design a pair
$(H,L)$ that renders an arbitrary mixed state $\rho$ GAS.  This problem
of GAS has been considered before for pure states, i.e., rank-$1$
projectors~\cite{wang-wiseman,ticozzi-QDS,ticozzi-attractive}.
Stabilizing pure states is an important task but it is not always
possible to stabilize a desired pure state.  However, there always
 exists a mixed state arbitrarily close to the target
state that can be stabilized, hence attaining {\em practical stability},
that is, stability of a neighborhood of the target state. Mixed-state
stabilization was considered in~\cite{schirmer-wang} but without a
constructive procedure to find the required dynamics $(H,L)$.  The $L$
\SGS{and $H$} we explicitly construct in this Note \SGS{are in a simple
form in an eigenbasis of $\rho$, tri- and quintdiagonal, respectively,
with off-diagonal elements determined by the spectrum of $\rho$.  The
diagonal elements $a_n$ and $h_n$ of $L$ and $H$ are free variables.
Further analysis shows that $\rho$ is GAS for most all choices of the
diagaonal elements $a_n$ and $h_n$, and we give explicit conditions on
$h_n$ to guarantee GAS of $\rho$.}

In Section \ref{qubit}, we illustrate by an example how the Lindblad
generator $L$ obtained this way can be implemented by reservoir
engineering via direct (Markovian) feedback~\cite{wiseman-feedback,
wang-wiseman,vitali-optomechanical, barchielli-gregoratti}, where the
controlled dynamics has the form of a Wiseman-Milburn Markovian
\emph{Feedback Master Equation} (FME)\cite{wiseman-feedback,
wiseman-milburn}:
\begin{equation}
\label{eq:FME}
 \dot{\rho}_t= -i[H+ H_C+\tfrac{1}{2}(FM+M^\dag F),\,\rho_t]
               +\D(M-iF,\rho_t)+\D(L_0,\rho_t).
\end{equation}
The drift Hamiltonian $H\in\HH(\H),$ the measurement operator
$M \in\BB(\H)$ and the noise operator $L_0$ are assumed to be
given, while the open-loop and the feedback Hamiltonians
$H_C, F\in\HH(\H)$ are our control parameters.

\section{Main Results}

\subsection{Preliminaries and design assumptions}
\label{preliminaries}

We recall two results on subspace stabilization that have been derived
in \cite{ticozzi-attractive,schirmer-wang}, which will be used in the
rest of the paper.

\begin{theorem}
[Feedback-attractive subspaces~\cite{ticozzi-attractive}]
\label{T-V-feedback} 
Let $\H_I=\H_S\oplus\H_R$ with $\Pi_S$ being the orthogonal projection
on $\H_S.$ If we can freely choose both the open-loop and the feedback
Hamiltonian then, for any measurement operator $M$, there exist a
feedback Hamiltonian $F$ and a Hamiltonian compensation $H_c$ that make
the subsystem supported by $\H_S$ invariant and attractive for the FME
\eqref{eq:FME} iff
\begin{equation}
 \label{stabilize}
 [\Pi_S,(M+M^\dag)] \neq 0.
\end{equation}
\end{theorem}
A constructive proof is provided in \cite{ticozzi-attractive}.  We will
also make use of the following characterization of the global
attractivity of a state:
\begin{theorem}
[Uniqueness equivalent to GAS \cite{schirmer-wang}]
\label{unique-attractive} 
A steady state of \eqref{gen} is GAS if and only if it is unique.
\end{theorem}

Finally, we require a few basic observations and a simple lemma.
Consider an orthogonal decomposition $\H=\H_S\oplus\H_R$ with
$d_s=\dim(\H_S),$ $d_r=\dim(\H_R)$, and let
$\{\ket{\phi_j^{S}}\}_{j=1}^{d_s},\{\ket{\phi_l^{R}}\}_{l=1}^{d_r}$
be orthonormal bases for $\H_{S},\,\H_R,$ respectively.  The basis
\begin{equation*}
  \{\ket{\varphi_m}\}
 =\{\ket{\phi_j^{S}}\}_{j=1}^{d_s}\cup\{\ket{\phi_l^{R}}\}_{l=1}^{d_r},
\end{equation*}
induces a block decomposition for matrices representing operators
acting on $\H$:
\begin{equation}
 \label{eq:blocks} 
  X=\left[\begin{array}{c|c} 
          X_S & X_P \\\hline X_Q &  X_R 
          \end{array}\right],
\end{equation}
and we have the following:
\begin{lemma}
 Assume $\rho=\left[\begin{smallmatrix}
        \rho_S          & 0                     \\
        0               & \rho_R
        \end{smallmatrix}\right],$
divided in blocks accordingly to some orthogonal Hilbert space
decomposition $\H=\H_S\oplus\H_R.$ Then $\rho$ is invariant for the
dynamics \eqref{gen} if and only if:
\begin{subequations}
\begin{align}
0=& -i[H_S,\rho_S]+L_S\rho_S L_S^\dag
    -\tfrac{1}{2}\{L_S^\dag L_S,\rho_S\}+L_P\rho_R L_P^\dag\nonumber\\
  & -\tfrac{1}{2}\{L_Q^\dag L_Q,\rho_S\}
    \label{condS}\\
0=&-i(H_P\rho_R-\rho_S H_P)+L_S\rho_S L_Q^\dag
    -\tfrac{1}{2}\rho_S(L_S^\dag L_P+L_Q^\dag L_R) \nonumber\\
  & +L_P\rho_R L_R^\dag
    -\tfrac{1}{2}( L_S^\dag L_P + L_Q^\dag L_R)\rho_R\label{condP}\\
0=&-i[H_R,\rho_R]+L_R\rho_R L_R^\dag
    -\tfrac{1}{2}\{L_R^\dag L_R,\rho_R\}+L_Q\rho_S L_Q^\dag\nonumber\\
  & -\tfrac{1}{2}\{L_P^\dag L_P,\rho_R\}
\label{condR}
\end{align}
\end{subequations}
\end{lemma}

\begin{IEEEproof}
By direct computation of the generator \eqref{gen} one finds its $S$,
$P$ and $R$-blocks to be the l.h.s. of \eqref{condS}-\eqref{condR},
respectively.  A given state $\rho$ is stationary if and only if
$\mathcal{L}(\rho)=0,$ and hence if and only if its blocks are all zero.
\end{IEEEproof}

\noindent We know from \cite{ticozzi-QDS, ticozzi-attractive} that for
$\rho$ to be invariant, its support $\H_\rho$ must be an invariant
subspace, and using the constructive procedure used in the proof of
Theorem \ref{T-V-feedback}, we can construct an block-upper-triangular
$L$ that stabilizes the $\H_\rho$ subspace.  There are many possible
choices to do that, e.g.
\begin{equation*}
 L= \left[ \begin{array}{c|c}
     L_\rho & L_{\rho,P} \\\hline
     0 & L_{\rho,R} \\
     \end{array}\right]\;
\end{equation*}
with blocks
\begin{equation*}
     L_{\rho,P}=
     \begin{bmatrix}
      0 & 0 & \cdots & 0 \\
      \vdots & 0 & \cdots & 0 \\
      \ell_1 & 0 & \cdots & 0 \\
     \end{bmatrix},\;
     L_{\rho,R}=
     \begin{bmatrix}
      0 & \ell_2 & 0 & 0 \\
      0 & 0 & \ell_3 & \ddots \\
      \vdots &  & \ddots & \ddots \\
\end{bmatrix}
\end{equation*}
with $\ell_1,\ell_2,\ldots \neq 0$.  Therefore, we can focus on the
dynamics restricted to the invariant support $\H_\rho$, and restrict
our attention here to full-rank states $\rho=\diag (p_1,\ldots,p_N)$
with $p_1,\ldots,p_{N}>0$.  To develop a constructive procedure to build
a stabilizing pair $(H,L)$ with a simple structure we make a series of
assumptions and design choices:

{\em Assumption 1.}  \emph{The \emph{spectrum} of $\rho$ is
{non-degenerate} (generic case).}

\noindent A state with non-degerate spectrum can be chosen
\emph{arbitrarily close to any state}.  Without loss of generality, we
can choose a basis such that $\rho$ is diagonal
$\rho=\diag(p_1,\ldots,p_N)$ with $p_1>\ldots>p_{N}>0$. This assumption
is instrumental for the construction of $L$ but can actually be relaxed,
as we remark after the Theorem \ref{main}.  Consider the decomposition
$\H=\H_S\oplus\H_R,$ with $\dim(\H_R)=1,$ such that the corresponding
block decomposition for $\rho$ is 
$\rho=\left[\begin{smallmatrix} \rho_S & 0 \\ 0 & \rho_R \\
\end{smallmatrix}\right],$ and divide accordingly $H, L$.

{\em Assumption 2.}  \emph{$\rho_S$ satisfies
\begin{equation}
 \label{hypS}
 -i[H_S,\rho_S]+L_S\rho_S L_S^\dag
 -\tfrac{1}{2}\{L_S^\dag L_S,\rho_S\}=0.
\end{equation}}

\noindent Condition \eqref{hypS} is clearly not satisfied in general,
but it will be ensured at each step of the iterative procedure outlined
in the next Section. Given this, conditions \eqref{condS}-\eqref{condR}
can be rewritten as:
\begin{equation}
\label{syst} \left\{
\begin{array}{l}
  0=L_Q\rho_S L_Q^\dag - \rho_R L_P^\dag L_P\\
  0=L_P\rho_R L_P^\dag - \tfrac{1}{2}\{L_Q^\dag L_Q,\rho_S\}.
\end{array}
\right.
\end{equation}
Since we assumed $\H_R$ to be one-dimensional, $L_P$ and $L_Q^\dag$ are
both $n-1$ dimensional vectors. Call $\Pi_P=L_P L_P^\dag/(L_P^\dag L_P),
\Pi_Q=L_Q^\dag L_Q/(L_Q L_Q^\dag).$ Then the second equation in
\eqref{syst} reads:
\begin{equation}
\label{Pis}
 \rho_R(L_P^\dag L_P) \Pi_P 
 =(L_Q L_Q^\dag)\tfrac{1}{2}(\Pi_Q\rho_S+\rho_S\Pi_Q).
\end{equation}

{\em Assumption 3.}  \emph{Choose $\Pi_Q$ to be the orthogonal projector
on the eigenspace corresponding to the smallest eigenvalue of $\rho_S$}.

\noindent Choose an ordered spectral decomposition with rank-one
projectors, $\rho=\sum_{j=1}^{N-1}p_i\Pi_{S,i}+\rho_R\Pi_R,$ where $p_i>
p_{i+1}.$ As $\rho$ is in diagonal form this means that we are choosing:
\begin{equation*}
  L_Q= \begin{bmatrix}0&\ldots &0 & \ell_Q \end{bmatrix}.
\end{equation*}
We thus get:
\begin{equation}
\label{key}
 \rho_R(L_P^\dag L_P) \Pi_P =(L_Q L_Q^\dag)p_{N-1}\Pi_Q,
\end{equation}
which can be satisfied if and only if $\Pi_P=\Pi_Q$ and $\rho_R(L_P^\dag
L_P)=(L_Q L_Q^\dag)p_{N-1}.$ We can choose $L_P^\dag L_P=p_{N-1},$
obtaining $(L_Q L_Q^\dag)={p_N}.$ Hence $\ell_Q$ can be in particular
chosen to be real, $\ell_Q=\sqrt{p_N}$.  We have thus constructed a
Lindblad term of the form:
\begin{equation*}
 L = \left[
      \begin{array}{c|c}
       L_S & \begin{array}{c} 
	     0\\ \vdots\\ 0\\ \sqrt{p_{N-1}}
	     \end{array}\\ \hline
      \begin{array}{cccc} 
      0& \cdots & 0 & \sqrt{p_{N}}
      \end{array} & L_R\\
      \end{array}
\right].
\end{equation*}


As $\rho_R$ is a scalar, we can rewrite \eqref{condP} as:
\begin{equation}
\label{condHp}
   -i(\rho_R I_S + \rho_S)H_P +K=0,
\end{equation}
where 
$K = -\tfrac{1}{2}( L_PL_R^\dag + L_S^\dag L_P + L_Q^\dag L_R)\rho_R 
     -\tfrac{1}{2}\rho_S(L_S^\dag L_P + L_Q^\dag L_R)+L_S\rho_SL_Q^\dag.$
Once $\rho_S,\rho_R,L_S,L_R,L_P,L_Q$ are fixed, \eqref{condHp} is a
linear system in $H_P$ which always admits a unique solution
$H_p=i(\rho_R I_S + \rho_S)^{-1}K,$ which is a necessary condition for
the invariance of $\rho$.  We are now in a position to \SGS{present an
inductive procedure for constructing stabilizing generators.}

\subsection{Constructive algorithm and proof of uniqueness}

We start by constructing a stabilizing generator for the two-level case.
Let $\rho=\left[\begin{smallmatrix} p_1 & 0 \\ 0 & p_2
\end{smallmatrix}\right]$ with $p_1> p_2>0.$  Note that in the
reasoning of the previous Section it is not necessary to impose
$\tr(\rho)=1.$ We will make use of this fact in extending the
procedure to the $n$-dimensional case.

Given the previous observations, we can render the given $\rho$
invariant for the dynamics if we can enact dissipation driven by a
Lindblad operator and a Hamiltonian of the form
\begin{equation}
\label{2level}
  L= \begin{bmatrix} a_1 & \sqrt{p_{1}} \\ \sqrt{p_{2}} & a_2 \end{bmatrix}
\end{equation}
and a $H$ such that its off-diagonal elements satisfy
\eqref{condHp}.


In the $N$-level case let $\rho=\diag(p_1,\ldots ,p_N)$.  We can iterate
our procedure by induction on the dimension of $\H_S.$ We have just
found the $2\times 2$ upper-left blocks of $L$ and $H$ such that
$\rho^{(2)}=\diag(p_1,p_2)$ is stable and attractive for the dynamics
driven by the reduced matrices.  Assume that we have some $m\times m$
upper-left blocks of $L$ and $H$ such that
$\rho^{(m)}=\diag(p_1,\ldots,p_m)$ is invariant for the reduced
dynamics.  This is exactly Assumption 2 above.  Let $\H_S^{(m)}=\supp
(\sum_{j=1}^{m}p_j\Pi_j),$ $\H_R^{(m)}=\supp(p_{m+1}\Pi_{m+1}).$ If we
want to stabilize $\rho^{(m+1)}=\diag(p_1,\ldots,p_{m+1})$ for the
dynamics restricted to $\H_S^{(m)}\oplus\H_R^{(m)}$, we can then
proceed building $L_Q^{(m)},L_P^{(m)}$ using Design Choice 1
above. Design Choice 2 lets us compute the off-diagonal terms of the
Hamiltonian, while we can pick $H_R$ to assume any value, since it does
not enter the procedure.

By iterating until $m=N,$ we obtain a tridiagonal matrix $L$ with
$L_{n,n+1}=\sqrt{p}_n$, $L_{n+1,n}=\sqrt{p}_{n+1}$ and
$L_{n,n}=a_n$, i.e.,
\begin{equation}
 \label{eq:L}
 L = \begin{bmatrix}
      a_1 & \sqrt{p}_1 & 0 & \cdots & 0 & 0\\
      \sqrt{p}_2 & a_2 & \sqrt{p}_2 & \cdots & 0 & 0\\
      0 & \sqrt{p}_3 & a_3 & \cdots & 0 & 0\\
      0 & 0 & \ddots & \ddots & \ddots & \vdots\\
      0 & 0 & \cdots & \sqrt{p}_{N-1} & a_{N-1}  & \sqrt{p}_{N-1}\\
      0 & 0 & \cdots & 0 & \sqrt{p}_N & a_N
   \end{bmatrix}
\end{equation}
and $H$ becomes a quintdiagonal Hermitian matrix, i.e.,
\begin{equation}
 \label{eq:H}
 \begin{split}
 H_{nn}   & = h_n, \\
 H_{n,n+1}&=H_{n+1,n}^* = \frac{i}{2}
\frac{\sqrt{p}_n-\sqrt{p}_{n+1}}{\sqrt{p}_n+\sqrt{p}_{n+1}}
              (a_n \sqrt{p}_n+ a_{n+1}\sqrt{p}_{n+1})\\
 H_{n,n+2}& =H_{n+2,n}^* = -\frac{i}{2} {p}_{n+1}
\frac{\sqrt{p}_n-\sqrt{p}_{n+2}}{\sqrt{p}_n+\sqrt{p}_{n+2}},
\end{split}
\end{equation}

\begin{theorem}
\label{main} If $(H,L)$ are chosen as in (\ref{eq:L}--\ref{eq:H}) then
$\rho=\diag({p}_1,\ldots,{p}_N)$ is a stationary state of the LME
$\dot\rho(t)=-i[H,\rho]+\D(L,\rho).$ \SGS{$\rho$ is GAS for most choices of the diagonal elements $\vec{a}$
and $\vec{h}$ of $L$ and $H$, respectively; in particular there exists $M_0\geq 0$ so that $\rho$ is GAS for $\vec{h}=(M,0,\ldots,0)$ for all $M>M_0.$
}
\end{theorem}

\begin{IEEEproof}
Given the generator, we can verify by direct calculation that $\rho$ is
a steady state.  Setting $B=\D(L,\rho)$, direct calculation shows that
$B_{m,n}=0$ except for
\begin{align*}
 B_{n,n+1} = B_{n+1,n}
 &=-\frac{1}{2} (\sqrt{p}_n-\sqrt{p}_{n+1})^2
    (d_n \sqrt{p}_n + d_{n+1}\sqrt{p}_{n+1}),\\
 B_{n,n+2} = B_{n+2,n}
 &=-\frac{1}{2} {p}_{n+1}(\sqrt{p}_n-\sqrt{p}_{n+1})^2,
\end{align*}
and setting $A=i[H,\rho]$ shows that the Hamiltonian term exactly
cancels the non-zero elements of $B$, i.e., $-A+B=0$.  Thus $\rho$ is a
steady state of the system.

To show how to make $\rho$ the unique, and hence attractive, stationary
state, assume that $\rho'$ is another stationary state in the support of
$\rho$.  Let $\X=\{X|X=x\rho+y\rho',\,x,y\in\RR\}.$ Then any state in
$\X\cap \DD(\H)$ is also stationary.  Since $\X$ is unbounded while
$\DD(\H)$ is compact, there must be a stationary state $\rho_1$ at the
boundary of $\DD(\H)$, i.e., with rank strictly less than $\rho$.  Then
the support of $\rho_1$, $\H_1=\supp(\rho_1)$, must be invariant
\cite{ticozzi-attractive}, and $H,L$ must exhibit the following
block-decompositions with respect to the orthogonal decomposition
$\H=\H_1\oplus\H_2:$
 \begin{equation*}
\label{eq:ss_bd3}
  H = \begin{bmatrix}
       H_{11} & H_{12} \\
       H_{12}^\dag & H_{22} \\
      \end{bmatrix}, \quad
  L = \begin{bmatrix}
        L_{11} & L_{12} \\
             0 & L_{22} \\
        \end{bmatrix},
\end{equation*}
with $H_{12}=-i\frac{1}{2} L_{11}^\dag L_{12}.$ If $L_{12}\neq 0$ then
it is straightforward to show that $\Tr(\Pi_2\rho),$ where $\Pi_2$ the
orthogonal projection onto $\H_2$, is strictly
decreasing~\cite{ticozzi-QDS}, and thus $\H_2$ is not invariant.  Since
$\rho$ is also stationary, and is a full-rank state, this is not
possible and we must therefore have $L_{12}=0$ and thus $H_{12}=0$.
This implies that $H$ and $L$ are block-diagonal.  Both $\H_1$ and
$\H_2$ must contain at least one eigenvector of $L,$ say $\vec{v}_1$ and
$\vec{v}_2.$  Orthogonality of the subspaces $\H_1$ and $\H_2$ along with the block-decomposition of $H$ and $L$ imply that any pair of vectors
$\vec{v}_1\in\H_1$, $\vec{v}_2\in\H_2$ must satisfy:
\begin{equation}
\label{eqs}
   \vec{v}_2^\dag L \vec{v}_1 
 = \vec{v}_2^\dag H \vec{v}_1 = \vec{v}_2^\dag \vec{v}_1 = 0.
\end{equation}
These are {\em necessary condition} the existence of another stationary
state, and hence for making $\rho$ not attractive.

From the results in the appendix (Theorem \ref{thm:NOE}), a tridiagonal
$L$ in the form (\ref{eq:L}) has $n$ distinct eigenvalues corresponding
to $n$ eigenvectors $\vec{v}_k$ with real entries and the first all
different from zero.  Without loss of generality, we can assume that the
first element of each (unnormalized) eigenvector $\vec{v}_j$ equals $1$
for all $j.$ If the eigenvectors are mutually non-orthogonal, i.e.,
$\vec{v}_k^T\vec{v}_\ell\neq 0$ for all $k,\ell$ then the third equality in
\eqref{eqs} is automatically violated, and hence $\rho$ must be the
unique stationary state.  This condition will almost always be satisfied
in practice, and it is easy to check that it always true when $N=2$ (see
appendix).  However, even if $L$ has orthogonal eigenvectors we can use
our freedom of choice in the diagonal elements $h_{nn}$ of $H$ to render
$\rho$ the unique stationary state.

Assume $\vec{v}_j^\dag \vec{v}_k = 0$ for some pair of eigenvectors of
$L$.  Let $H_0$ be the Hamiltonian corresponding to $\vec{h}=\vec{0}$
and $M_0=\max_{j,k}|\vec{v}_j^\dag H_0 \vec{v}_k|$.  Choose $M>M_0$ and
let $H$ be the Hamiltonian corresponding to $\vec{h}=(M,0,\ldots)$,
$H=H_0+\diag(M,0,\ldots,0)$.  Recalling that the first component of
$\vec{v}_j$ is $1$ for all $j$, shows that
\begin{equation*}
  \vec{v}_j^\dag H \vec{v}_k = M + \vec{v}_j^\dag H_0 \vec{v}_k > 0,
\end{equation*}
and therefore the second equality in \eqref{eqs} is violated, and $\rho$
is the unique stationary state.
\end{IEEEproof}

{\em Remark:} Theorem \ref{main}--\ref{thm:NOE} further shows that
{Assumption 1} on the spectrum of $\rho$ can be relaxed, since the
fact that the spectrum is non-degenerate plays no role in the proof. The
construction is effective for {\em any} full rank state on the desired
support.

\section{Feedback stabilization of encoded qubit states}
\label{qubit}

Consider a system whose evolution is governed by the FME~(\ref{eq:FME})
with $H_0=-H_C'$, $M=M^\dag,$ and $L$ admitting an eigenspace $\H_S$ of
dimension 2 for some eigenvalue $\lambda_S$.  Assume we can switch off
the measurement and the feedback Hamiltonian, $F=M=0.$ With this choice,
\eqref{eq:FME} admits a two-dimensional \emph{noiseless (or
decoherence-free) subspace}~\cite{lidar-DFS,
lidar-initializationDFS,ticozzi-QDS}, which can be effectively used to
encode a quantum bit protected from noise. We now face the problem of
{\em initializing} the quantum state {\em inside the DFS}: we thus wish
to construct $H_C$, $F$ and $M,$ such that a given state $\rho$ \emph{of
the encoded qubit} is GAS on the full Hilbert space.  Setting
$\H=\H_S\oplus\H_R$ and choosing an appropriate basis for $\H_S$,
the encoded state to be stabilized takes the form
\begin{equation*}
 \rho  = \begin{bmatrix} \rho_S & 0\\  0 & 0\end{bmatrix}, \quad
 \rho_S= \begin{bmatrix}  p_1 & 0\\ 0 & p_2 \end{bmatrix}.
\end{equation*}
The dynamical generators can be partitioned accordingly,
\begin{gather*}
 L  = \begin{bmatrix} \lambda_M I_2 & L_P\\ 0 & L_R \end{bmatrix}, \quad
 H_C= \begin{bmatrix} H_S & H_P\\ H_P^\dag & H_R \end{bmatrix}, \\
 F  = \begin{bmatrix} F_S & F_P\\ F_P^\dag & F_R \end{bmatrix}, \quad
 M  = \begin{bmatrix} M_S & M_P\\ M_P^\dag & M_R \end{bmatrix}
\end{gather*}
where $I_2$ is the $2\times 2$ identity matrix.  We compensate the
feedback-correction to the Hamiltonian by choosing
$H_C=H_C'-\frac{1}{2}(FM+M^\dag F)$.  For $L_P\neq0$ we use the
constructive algorithm described above to render $\H_S$ attractive
by choosing $F_P=-iM_P$, and constructing a
$H_R'$ for $H_C'$ so that no invariant state has support in $\H_R$
(see Theorem 12 in \cite{ticozzi-attractive}) and by imposing
$H'_P=0$.  For $L_P=0$ we need to choose an observable such that
$M_P\neq0$.  We are thus left with freedom on the choice of
$H_S,F_S,$ which can be now used to stabilize the desired $\rho_S$
in the {controlled invariant subspace} $\H_S.$  
Denote the elements
of the upper-left $2\times 2$ blocks as
\begin{equation}
  M_S= \begin{bmatrix} M_1 & M_3\\ M_3^\dag & M_2\end{bmatrix}, \quad
  F_S= \begin{bmatrix} F_1 & F_3\\ F_3^\dag & F_2\end{bmatrix}
\end{equation}
Set $F_{3}=-\frac{ik_M}{2}(\sqrt{p}_2-\sqrt{p}_{1})$.  If $M_3\neq 0$,
setting $k_M:=2M_3/(\sqrt{p}_1+\sqrt{p}_{2})$ shows that
$M_{3}=\frac{k_M}{2}(\sqrt{p}_1+\sqrt{p}_{2})$, i.e., up to the
multiplicative constant $k_M$ the two block matrices are such that
$M_S-iF_S$ is of the form \eqref{eq:L}, Theorem \ref{main} applies and
the desired state is GAS. Notice that if $M_3=0,$ and $M$ has
non-degenerate spectrum, it is easy to find another state $\rho'$,
arbitrarily close to $\rho,$ such that, in the basis in which $\rho'$ is
diagonal, we have $M_3'\neq 0.$ Thus we can attain practical
stabilization of any state of the encoded qubit.

\vspace{-5mm}
\section{Conclusions and outlook}

Efficient quantum state preparation is crucial to most of the physical
implementations of quantum information technologies.  Here we have shown
how \emph{quantum noise} can be designed to stabilize arbitrary
quantum states. The main interest in such a result is
motivated by direct feedback design for applications in quantum optical
and opto-mechanical systems and quantum information processing
applications. As an example we have demonstrated how to devise the
(open- and closed- loop) control Hamiltonians in order to asymptotically
stabilize a state of a qubit encoded in a noiseless subspace of a larger
system.  Further study is under way to address similar problems for
multiple qubits, in the presence of structural constraints on the
measurements, control and feedback operators, and to optimize the speed
of convergence to the target state.

\section*{Acknowledgments}

We thank Arieh Iserles, Lorenza Viola and Maria Jose Cantero for
interesting and fruitful discussions.

\appendix

\section{Eigenvalues and eigenvectors of tridiagonal matrices}

We collect here results about orthogonal polynomials and tridiagonal
matrices that are instrumental to the proof of Theorem~\ref{main}.
Details and (missing) proofs can be found in
\cite{Marcellan,Totik,Chihara}.

\begin{definition}
An orthogonal polynomial sequence $\{P_n(x)\}_{n=1}^\infty$ over
$[a,b]\subset\RR$ is an infinite sequence of real polynomials such
that $\langle P_j,P_k\rangle=0$ for any $j\ne k$ under some $L^2$
inner product $\langle P_j, P_k \rangle:=\int P_j(x) P_k(x) w(x) dx$
with a weighing function $w(x)$.
\end{definition}

\begin{theorem}
[Favard~\cite{Marcellan}] \label{thm:Favard} A monic polynomial
sequence $\{P_n(x)\}$ is an orthogonal polynomial sequence if it
satisfies a three-term recurrence relation
\begin{equation}
   P_{n+1}(x) = (x-b_n) P_n(x) - c_n P_{n-1}(x), n \ge 0
\end{equation}
with $\{b_n\}$, $\{c_n\}$ sequences of real numbers and $c_n>0$.
\end{theorem}

\begin{theorem}
[\cite{Totik}] \label{thm:MPS} If $\{P_n(x)\}$ is a sequence of
orthogonal polynomials then each $P_n(x)$ has $n$ distinct roots.
\end{theorem}

\begin{theorem}
\label{thm:Neig} A real-symmetric tridiagonal $N\times N$ matrix with
non-zero entries on the first sub/superdiagonal has $N$ distinct
eigenvalues. \end{theorem}

\begin{IEEEproof}
It suffices to note that the monic polynomial sequence $\{f_n\}$ with
$f_0(\lambda)=1$ and
\begin{equation}
  f_n(\lambda)= \det \begin{bmatrix}
   \lambda-\alpha_1 & \beta_1 & 0       & \ldots & 0 \\
   \beta_1 & \lambda-\alpha_2 & \beta_2 &        & \vdots \\
   0       &                  & \ddots  &        &            \\
   \vdots  & \vdots &         &         &        &            \\
           &        &         &         &       \beta_{n-1}\\
   0       & \cdots & 0       & \beta_{n-1} & \lambda-\alpha_n
\end{bmatrix}
\end{equation}
for $n=1,2,\ldots N$ satisfies the recurrence relation
\begin{equation}
\label{recurrence}
  f_{n+1}(\lambda) = (\lambda-\alpha_{n+1})f_n(\lambda)
                      -\beta_n^2 f_{n-1}(\lambda)
\end{equation}
and therefore is an orthogonal polynomial sequence by Favard's theorem
provided $\beta_n^2>0$.  Thus $f_n(\lambda)$ has $n$ distinct roots for
all $n$, and the matrix has $N$ distinct eigenvalues.\end{IEEEproof} The
expression for the corresponding (unnormalized) eigenvectors
$\tilde{\vec{v}}_k$ can then be obtained by straightforward calculation:
\begin{equation}
  \label{elements}
   \tilde{v}_{jk}:= \frac{f_{j-1}(\lambda_k)}{\beta_1\cdots\beta_{j-1}}.
\end{equation}

\begin{theorem}
\label{thm:NOE} 
Let $T$ be a real tridiagonal matrix.  If $T_{n,n+1}=\beta_n>0$ and
$T_{n+1,n}=\gamma_n>0$ then $T$ has $N$ distinct eigenvalues
$\lambda_n$.  
\end{theorem}

\begin{IEEEproof}
There always exists a diagonal matrix $D=\diag(d_n)$ such that $S =
D^{-1} T D$ is a symmetric tridiagonal matrix: the diagonal elements
must satisfy $d_{n+1}^2=d_n^2\gamma_n/\beta_n$ for all $n$, which shows
that $D$ is uniquely determined up to global factor. We fix $d_n=1$. The
off-diagonal elements of $S$ are $S_{n,n+1} = S_{n+1,n} =
\sqrt{\beta_n\gamma_n}$.  The matrix $S$ is real-symmetrix and hence
$S=VEV^\dag,$ with $E=\diag(\lambda_n).$ By the previous theorem the
eigenvalues $\lambda_n$ are real and distinct provided
$\beta_n\gamma_n>0$ for all $n$.  $V$ is a real-orthogonal matrix whose
columns are the normalized eigenvectors of $S$,
$\vec{v}_k=C_k^{-1}\tilde{\vec{v}}_k$ with $\vec{v}_{jk}$ as defined
in~(\ref{elements}) and $C_k=\norm{\tilde{\vec{v}_k}}$.  Since $T =
DSD^{-1} = D(VEV^T)D^{-1} = (DV)E(DV)^{-1},$ $T$ has the same
eigenvalues as $S,$ with eigenvectors $\vec{w}_k=D\vec{v}_k.$
\end{IEEEproof}

While it is not strictly needed for the result in this work, it is worth
noticing that for almost all choices of diagonal entries in $T,$ the
eigenvectors will be mutually non-orthogonal \SGS{unless $T$ is
symmetric.} For fixed off-diagonal elements $\beta_k$ and $\gamma_k$,
the eigenvalues $\lambda_k$ and corresponding eigenvectors $\vec{v}_k$
of $S$ can be expressed explicitly in terms of the diagonal entries
$\vec{a}$.  Since $f_N(\lambda,\vec a)$ is an $N$th order polynomial in
$\lambda$ with $N$ distinct roots, by the implicit function theorem, all
roots $\lambda_k$ for $f_N(\lambda,\vec{a})=0$ can be expressed locally
as continuously-differentiable functions $\lambda_k(\vec{a})$ in some
open neighborhood $\N_{\vec{a}}$ of $\vec{a}$, and similarly for the
eigenvectors $\vec{v}_k$.  To determine whether $T(\aa)$ has orthogonal
eigenvectors consider the functions
\begin{equation}
   F_{k\ell}(\aa)
   = \sum_{n=1}^{N-1} d_n^2 b_{n-1}^2 
      f_{n-1}(\lambda_k(\aa)) f_{n-1}(\lambda_\ell(\aa))
\end{equation}
which correspond to the inner products of the unnormalized eigenvectors
of $T$ with $b_n=(\beta_{1}\ldots\beta_n\gamma_1\ldots\gamma_n)^{-1}$.  
By orthogonality of the eigenvectors of $S$ we have
\begin{equation}
   B_{k\ell} = \sum_{n=1}^{N-1} b_{n-1}^2 
        f_{n-1}(\lambda_k(\aa)) f_{n-1}(\lambda_\ell(\aa))
\end{equation}
for all $\aa$.  But $F_{k\ell}$ and $G_{k\ell}$ are polynomials in
$\lambda$ and $\aa$ and differ only by the coefficiently $d_n$.  Thus,
we will have $F_{k\ell}(\aa)\neq G_{k\ell}(\aa)$ for most $\aa$ unless
all $d_n$ are equal, \SGS{which implies $\beta_n=\gamma_n$ for all $n$
and $T$ symmetric.  (In our case this would only happen when all
populations $p_n$ are equal.)  By continuity of $F_{k\ell}(\aa)$ we can
conclude that in any neighborhood $\N_{\aa_0}$ of a bad point $\aa_0$
with $F_{k\ell}(\aa_0)=0$, there is an open (and thus positive-measure)
subset of points with $F_{k\ell}(\aa)\neq0$, provided $F_{k\ell}(\aa)$
is not constant on $\N_{\aa_0}$.}  This argument can be made stronger
and more rigorous if we assume $\frac{\d
F_{k\ell}(\aa)}{\d\alpha_m}(\aa_0)\neq0$ for some $\alpha_m$, i.e., that
not all partial derivatives at the bad point $\aa_0$ vanish, in which
case we can show that $F_{k\ell}(\aa)$ vanishes only on a measure-zero
subset of $\N_{\aa_0}$ \SGS{using the implicit function theorem.}

\end{document}